\documentclass{article}
\usepackage{amssymb,amsmath,latexsym,amscd,amsfonts}
\usepackage{graphics}
\usepackage{epsfig}

\newtheorem{thm}{Theorem}
\newtheorem{proposition}{Proposition}
\newtheorem{definition}{Definition}

\newtheorem{corol}{Corollary}

\newenvironment{proof}{{\noindent{\bf Proof:}}}{$\hfill\Box$}

\def\NP{{\sf{NP}}}

\def\NEXP{{\sf{NEXP}}}

\def\PSPACE{{\sf{PSPACE}}}

\def\QIP{{\sf{QIP}}}

\def\MIP{{\sf{MIP}}}

\def\Tr{{\textnormal{Tr}}}

\def\hpic #1 #2 {\mbox{$\begin{array}[c]{l}
\epsfig{file=#1,height=#2} \end{array}$}}
\def\vpic #1 #2 {\mbox{$\begin{array}[c]{l}
\epsfig{file=#1,width=#2}\end{array}$}}

\begin{document}
\title{A Lower Bound on the Value of Entangled Binary Games}

\author{
Salman Beigi\\ {\it \small Institute for Quantum Information}\\ {\it \small California Institute of
Technology}\\ {\it \small Pasadena, CA}
}

\date{}

\maketitle{}

\begin{abstract} A two-player one-round binary game consists of two cooperative players who each replies by one bit to a message that he receives privately; they win the game if both questions and answers satisfy some predetermined property. A game is called entangled if the players are allowed to share a priori entanglement. It is well-known that the maximum winning probability (value) of entangled XOR-games (binary games in which the predetermined property depends only on the XOR of the two output bits) can be computed by a semidefinite program. In this paper we extend this result in the following sense; if a binary game is uniform, meaning that in an optimal strategy the marginal distributions of the output of each player are uniform, then its entangled value can be efficiently computed by a semidefinite program. We also introduce a lower bound on the entangled value of a general two-player one-round game; this bound depends on the size of the output set of each player and can be computed by a semidefinite program. In particular, we show that if the game is binary, $\omega_q$ is its entangled value, and $\omega_{sdp}$ is the optimum value of the corresponding semidefinite program, then $0.68\,\omega_{sdp} < \omega_q \leq  \omega_{sdp}$.

\end{abstract}

\section{Introduction}

Consider a game with two players, Alice and Bob, in which they upon receiving questions $s$ and $t$, respectively, drawn from some probability distribution $\pi(s, t)$, output $a$ and $b$. They win the game if the value of a predetermined Boolean function $V(a, b, s, t) = V(a, b \vert s, t)$ is $1$. Alice and Bob can agree on a strategy before starting the game, but no communication is allowed afterwards. Such a game is called a two-player one-round game. Finding the best strategy and the winning probability, called the {\it value}, of these games is the main concern of this paper.

Two-player one-round games have been introduced in order to study the power of multi-prover interactive proofs ($\MIP$). It is known that given an explicit description of an efficiently computable probability distribution $\pi$ and the Boolean function $V$, it is $\NP$-complete to decide whether the winning probability of the game is greater than a given bound or not. Indeed, this statement still holds even if Alice and Bob's answers are binary ($a, b\in \{0,1\}$), and the function $V(a, b\vert s, t)$ depends only on the XOR of their answers ($V(a, b\vert s, t)= V(a\oplus b\vert s, t)$). Moreover, by translating this result into the language of interactive proofs (which gives games of exponential size), the complexity class of problems that admit a two-prover one-round XOR proof system with completeness $\alpha=12/16-\epsilon$ and soundness $\beta=11/16+\epsilon$, for every $0 < \epsilon <1/16$, is equal to $\NEXP$ ($\oplus\MIP_{\alpha, \beta}[2,1] = \NEXP$) \cite{CHTW, Hastad}.

Taking quantum physics into account, Alice and Bob may increase their winning probability by sharing an entangled state; they determine $a$ and $b$ according to local measurements of the shared state in some basis depending on their questions (Alice's basis depends on $s$ and Bob's basis depends on $t$). Such a game is called an entangled game.

Estimating the value of entangled games is a hard problem in general since there is no known bound on the amount of entanglement that should be shared between the two players in order to reach the optimal strategy. We do not know even if some finite amount of entanglement is enough or not (see \cite{dltw, npa,vertesi,grothen} for some results in this direction). Also, no relation has been established between $\MIP^{\ast}_{\alpha, \beta}[2,1]$ (two-prover one-round interactive proofs with shared entanglement, and with completeness $\alpha$ and soundness $\beta$; $\ast$ denotes the presence of entanglement) {\it with constant gap} and $\NEXP$.\footnote{Ito, Kobayashi, and Matsumoto in \cite{ito} have shown that $\NEXP$ is contained in $\bigcup \MIP^{\ast}_{1, 1-2^{-p(n)}}[2,1]$, where the union is taken over all polynomials $p(n)$; in their protocol for $\NEXP$ the gap is exponentially small, but the output of each prover is only two bits. In the same paper they have proved that $\PSPACE$ is in $\MIP^{\ast}_{1, 2^{-p(n)}}[2,1]$ for every polynomial $p(n)$. For other results in this direction see \cite{19} and \cite{20}.} However, in some special cases the value of entangled games can be efficiently approximated.

The winning probability of XOR-games can be expressed as a semidefinite program (SDP) and be computed efficiently \cite{Tsirelson, CHTW}. Indeed, this SDP can be solved in $\PSPACE$; Wehner \cite{ste} has shown that $\oplus \MIP^{\ast}_{\alpha, \beta}[2,1]$ is in $\QIP_{\alpha, \beta}(2)$ (two-message quantum interactive proofs) which combined with the celebrated result of Jain, Ji, Upadhyay and Watrous \cite{qip-pspace} (see also \cite{qip2-pspace}) that $\QIP=\PSPACE$ gives $\oplus \MIP^{\ast}_{\alpha, \beta}[2,1]\subseteq \PSPACE$, for every $\alpha$ and $\beta$ with an inverse polynomial gap. Regarding entangled XOR-games we also know that $\NP\subseteq \oplus\MIP^{\ast}_{1-\epsilon, 1/2+\epsilon}[2,1]$, for every $0< \epsilon <1/2$, \cite{CGJ}.

We can also estimate the value of entangled {\it unique games}. A game is called unique if for every $s$ and $t$, and any given answer $a$ from Alice there exists a unique $b$ such that $V(a, b\vert s, t)=1$ and vice versa (for every $b$ there exists a unique $a$ with $V(a, b\vert s, t)=1$). Kempe, Regev, and Toner in \cite{unique} have shown that the corresponding SDP to a unique game provides us with a close approximation of the value of the game; if the value of the game is $1-\epsilon$, it efficiently obtains a strategy with winning probability at least $1-6\epsilon$. They also have shown that this estimation can be improved if the game is {\it uniform} as well.\footnote{A game is called uniform if there exists an optimal strategy such that for every questions $s$ and $t$, the marginal distributions of Alice and Bob's output are uniform.}

For a general game, one can define a hierarchy of semidefinite programs that ignoring some technicalities, in the limit gives the exact value of an entangled game \cite{npa, dltw}. However, we do not know up to which level of the hierarchy one should go in order to reach a fair approximation of the optimal value. \\

\noindent{\bf Our contribution:} The focus of this paper is estimating the value of entangled games with a small set of outputs. We consider binary games ($a, b\in \{0, 1\}$) which are not necessarily XOR-games. It is known that entangled Alice and Bob can win a binary game with probability $1$ if and only if they can always win without entanglement \cite{CHTW}. But in general no relation is known between the classical value and quantum value of binary games. In this paper we show that if a binary game is uniform, its entangled value can be efficiently computed by an SDP (see Theorem \ref{thm:uniform}). By generalizing the ideas of this result, we also provide a lower bound on the value of a general entangled game as follows.

We associate an SDP (see Section \ref{sec:sdp}) to a game $G$ with the optimum value $\omega_{sdp}(G)$. Due to the description of the SDP it is easy to see that $\omega_{sdp}(G)$ is an upper bound for $\omega_q(G)$, the entangled value of $G$. Here we prove the lower bound
\begin{equation}
\omega_q(G)\geq c_k \omega_{sdp}(G),
\end{equation}
where $k$ is the size of the output set of Alice and Bob ($a, b\in \{1, \dots, k\}$), and  $c_2\approx 0.68$ and $c_k=1/(k-1)$ for $k\geq 3$ (see Theorem \ref{thm:general} below).

These results are based on the techniques of Tsirelson \cite{Tsirelson} who for proving that the value of XOR-games can be computed by an SDP, provides an embedding of unit vectors into the space of observables (hermitian matrices with $+1, -1$ eigenvalues) that preserves the inner product. The point is that the entangled value of XOR-games can be expressed in terms of observables, and then this embedding provides a way to construct a strategy from a solution of the SDP. In general, however, the value of an entangled game is given by an optimization problem over POVM measurements, so we should directly deal with POVM elements instead of observables. Here based on a similar construction as in \cite{Tsirelson}, we provide an embedding of a real vector space into the space of positive semidefinite matrices which {\it approximately} preserves the inner product. As a result, we obtain an estimation of the entangled value of games based on their SDP relaxation.

\section{Entangled games and their SDP relaxation}\label{sec:sdp}

\begin{definition} A two-player one-round game $G$ consists of a probability distribution $\pi$ on a set of the form $S\times T$, and a function $V: A\times B\times S\times T \rightarrow \{0,1\}$, where $A$ and $B$ are two given sets. The game starts by choosing $(s,t)\in S\times T$ according to distribution $\pi$, sending $s$ to the first player (called Alice) and $t$ to the second player (called Bob); then Alice and Bob output $a\in A$ and $b\in B$, respectively. They win the game if $V(a, b, s, t)=V(a, b\vert s, t)=1$.
\end{definition}

The value of a game is the maximum winning probability of Alice and Bob over all strategies. This optimum value in the classical case is denoted by $\omega_c(G)$. In the quantum case where Alice and Bob can use quantum resources the game is called an entangled game and its maximum winning probability is denoted by $\omega_q(G)$.

A general strategy of Alice and Bob for playing an entangled game has the following structure. They share a bipartite state $\vert \phi\rangle$ and depending on their questions, locally measure it to decide about their output messages. Without loss of generality we may assume that the measurements are projective, so Alice upon receiving $s\in S$ performs the projective measurement $\{P_1^{s}, \dots, P_k^s\}$, and similarly let $\{Q_1^t, \dots, Q_k^t\}$ be the projective measurement of Bob corresponding to message $t\in T$. (Here we assume that $A=B=\{1, \dots, k\}$.) Thus the probability that Alice and Bob output $a\in A$ and $b\in B$ is $\langle \phi\vert P_a^s\otimes Q_b^t\vert \phi\rangle$, and the winning probability of this strategy is equal to
\begin{equation}\label{eq:1}
\sum_{s,t} \pi(s, t) \sum_{a,b} V(a, b\vert s,t) \langle \phi\vert P_a^s\otimes Q_b^t\vert \phi\rangle.
\end{equation}
Then, $\omega_q(G)$ is equal to the supremum value of this expression over all choices of $\vert \phi\rangle$, $\{P_a^s\}$ and $\{Q_b^t\}$.

Associated to any entangled game there is a corresponding SDP which provides an upper bound on the value of the game. For any state $\vert \phi\rangle$, and projective measurements $\{P_a^s\}$ and $\{Q_b^t\}$ define the vectors
\begin{equation}
\vert v_a^s\rangle = P_a^s\otimes I \vert \phi\rangle,
\end{equation}
and
\begin{equation}
\vert w_b^t\rangle = I\otimes Q_b^t \vert \phi\rangle.
\end{equation}
Therefore,
\begin{equation}
\sum_{s,t} \pi(s, t) \sum_{a,b} V(a, b\vert s,t) \langle \phi\vert P_a^s\otimes Q_b^t\vert \phi\rangle = \sum_{s,t} \pi(s, t) \sum_{a,b} V(a, b\vert s,t)  \langle v_a^s\vert w_b^t\rangle.
\end{equation}
We conclude that the maximum value of the right hand side of this equation (over all vectors $\vert v_a^s\rangle$ and $\vert w_b^t\rangle$) is an upper bound for $\omega_q(G)$.

Since $\{P_a^s\}$ and $\{Q_b^t\}$ are measurement operators the vectors $\vert v_a^s\rangle$ and $\vert w_b^t\rangle$ satisfy some constraints. First, since $\{P_a^s\}$ is a projective measurement $P_a^sP_{a'}^s=0$, for any $s$ and $a\neq a'$, and then $\langle v_a^{s}\vert v_{a'}^s\rangle = \langle \phi\vert P_a^sP_{a'}^s\otimes I\vert \phi\rangle =0 $. Similarly, $\langle w_b^{t}\vert v_{b'}^t\rangle=0$ for any $t$ and $b\neq b'$. Also, since for any $s$, $\sum_a P_a^s=I$ we have
\begin{equation}
\sum_a \vert v_a^s\rangle = \sum_a P_a^s\otimes I\vert \phi\rangle = \vert \phi\rangle,
\end{equation}
and similarly $\sum_b \vert w_b^t\rangle = \vert \phi\rangle$. Thus we obtain the following SDP
\begin{alignat}{2}
    \text{Maximize:} \quad & \sum_{s,t} \pi(s, t) \sum_{a,b} V(a, b\vert s,t)  \langle v_a^s\vert w_b^t\rangle \nonumber \\
    \text{Subject to:} \quad
    & \langle z\vert z\rangle =1, \nonumber\\
    & \vert v_1^s\rangle + \cdots + \vert v_k^s\rangle = \vert z\rangle, \quad & \forall s\in S\nonumber\\
    & \vert w_1^t\rangle + \cdots + \vert w_k^t\rangle = \vert z\rangle, \quad & \forall t\in T\nonumber\\
    & \langle v_a^s\vert v_{a'}^s\rangle =0, \quad & \forall s\in S, a\neq a'\in A \nonumber\\
    & \langle w_b^t\vert w_{b'}^t\rangle =0, \quad & \forall t\in T, b\neq b'\in B \nonumber\\
    & \langle v_a^s\vert w_b^t\rangle \geq 0, \quad & \forall s\in S, t\in T, a\in A, b\in B. \label{eq:sdp}
\end{alignat}
Observe that although the second and third constraints are not in terms of inner products, they simply can be written in that form (see \cite{unique}). Here this SDP is considered over real numbers; that is, the vectors $\vert v_a^s\rangle, \vert w_b^t\rangle$, and $\vert z\rangle$ have real coordinates.

Letting $\omega_{sdp}(G)$ be the optimum value of the above SDP, we conclude that
\begin{equation}
\omega_{q}(G) \leq \omega_{sdp}(G).
\end{equation}

\section{Uniform binary games}\label{sec:uniform}

An entangled game is called uniform if there exists an optimal strategy in which the marginal distributions of Alice and Bob's output are uniform, i.e., for any $s$ and $t$, the probability that Alice (Bob) outputs $a\in \{1, \dots, k\}$ ($b\in \{1, \dots, k\}$) is $1/k$. Using our notation, this is equivalent to $\langle \phi\vert P_a^s\otimes I\vert \phi\rangle = \langle \phi\vert I\otimes Q_b^t\vert \phi\rangle =1/k$ for any $s, t, a$ and $b$. For example, any XOR-game is uniform because Alice and Bob can share a random bit, play the optimal strategy, and then both flip their outputs depending on the random bit; since $a+b \mod 2$ does not change if one flips both $a$ and $b$, the winning probability of this new uniform strategy remains the same.

It is shown by Tsirelson \cite{Tsirelson} that for any XOR-game the upper bound $\omega_{sdp}(G)$ is equal to $\omega_q(G)$. Here, we extend this result and show that for any uniform binary game $G$, $\omega_{q}(G)$ can be computed by an SDP.

The corresponding SDP to a uniform binary game is slightly different from the one given in the previous section. Indeed, since the game is uniform for every $s$ and $a$ we have
\begin{equation}\label{eq:ex1}
\langle v_a^s\vert v_a^s\rangle = \langle \phi\vert P_a^s\otimes I\vert \phi\rangle = \frac{1}{2},
\end{equation}
and
similarly for every $t$ and $b$
\begin{equation}\label{eq:ex2}
\langle w_b^t\vert w_b^t\rangle = \frac{1}{2}.
\end{equation}
For a binary game $G$ we let $\omega_{sdp}^u(G)$ be the optimum value of the SDP (\ref{eq:sdp}) together with the extra constraints (\ref{eq:ex1}) and (\ref{eq:ex2}).

\begin{thm} \label{thm:uniform}
$\omega_{q}(G)= \omega_{sdp}^u(G)$, for every uniform binary game $G$.
\end{thm}

The proof of this theorem follows from almost the same steps as in the proof of Tsirelson for XOR-games. The only difference is that in XOR-games, $\omega_q(G)$ can be expressed in terms of observables  which does not hold for a general binary game. So instead of observables we directly work on projections and try to follow similar steps.\\

\begin{proof} $\omega_{sdp}^{u}(G)\geq \omega_q(G)$ is already discussed. For the other direction let the vectors $\vert z\rangle, \vert v_a^s\rangle, \vert w_b^t\rangle$, for all $s, t, a$ and $b$, denote the optimum point of the SDP (\ref{eq:sdp}) with the extra constraints (\ref{eq:ex1}) and (\ref{eq:ex2}). Assume that these vectors live in the $(m+1)$-dimensional real space. We show that there exists a strategy for Alice and Bob with winning probability
\begin{equation}
\omega_{sdp}^u(G) = \sum_{s,t} \pi(s, t) \sum_{a,b} V(a, b\vert s,t)  \langle v_a^s\vert w_b^t\rangle.
\end{equation}

Without loss of generality we may assume that\footnote{We can rotate all the vectors in the space; also note that $\vert z\rangle$ is a unit vector.}
\begin{equation}\label{eq:z}
\vert z\rangle = \left(
                   \begin{array}{c}
                     1 \\
                     0 \\
                     \vdots \\
                     0 \\
                   \end{array}
                 \right).
\end{equation}
Then for any $s$ since $\vert v_0^s\rangle + \vert v_1^s\rangle = \vert z\rangle$ we have
\begin{equation}\label{eq:vx}
\vert v_0^s\rangle = \left(
                   \begin{array}{c}
                     x_{s,0} \\
                     x_{s,1} \\
                     \vdots \\
                     x_{s,m} \\
                   \end{array}
                 \right), \hspace{.5in}
\vert v_1^s\rangle = \left(
                   \begin{array}{c}
                     1-x_{s,0} \\
                     -x_{s,1} \\
                     \vdots \\
                     -x_{s,m} \\
                   \end{array}
                 \right),
\end{equation}
for some real numbers $x_{s,0}, \dots, x_{s,m}$.
Since $\vert v_0^s\rangle$ and $\vert v_1^s\rangle$ are orthogonal,
\begin{equation}1/2=\langle v_0^s\vert v_0^s\rangle =  \langle z\vert v_0^s\rangle = x_{s,0}.
\end{equation}
Moreover, from $1/2=\langle v_0^s\vert v_0^s\rangle=\sum_{i=0}^{m} x_{s,i}^2$ we conclude that
\begin{equation}\label{eq:x1/4}
\sum_{i=1}^m x_{s,i}^2 = \frac{1}{4}.
\end{equation}
Similarly, for any $t$ we have
\begin{equation}\label{eq:wy}
\vert w_0^t\rangle = \left(
                   \begin{array}{c}
                     y_{t,0} \\
                     y_{t,1} \\
                     \vdots \\
                     y_{t,m} \\
                   \end{array}
                 \right), \hspace{.5in}
\vert w_1^t\rangle = \left(
                   \begin{array}{c}
                     1-y_{t,0} \\
                     -y_{t,1} \\
                     \vdots \\
                     -y_{t,m} \\
                   \end{array}
                 \right),
\end{equation}
where $y_{t,0}=1/2$ and $\sum_{i=1}^m y_{t,i}^2 = 1/4$.

Let $\sigma_1, \dots, \sigma_m$ be some $d\times d$ matrices such that $\sigma_i^2=I$, $\Tr\, \sigma_i =0$, $\sigma_i^{\dagger}=\sigma_i$, and for any $i\neq j$, $\sigma_i\sigma_j = -\sigma_j\sigma_i$. Such matrices can be found in dimension $d=2^m$ by taking 
\begin{align}\sigma_i= \sigma_z\otimes \cdots \otimes \sigma_z \otimes \sigma_x\otimes I \otimes\cdots \otimes I,
\end{align} 
consisting of the tensor product of Pauli operators $\sigma_z$, $\sigma_x$, and identity, where $\sigma_x$ is the $i$-th Pauli matrix in the tensor product. Also let $\sigma_0=I$, and for any $(m+1)$-dimensional vector define the matrix
\begin{equation}
\left(
  \begin{array}{c}
    u_0 \\
    u_1 \\
    \vdots \\
    u_m \\
  \end{array}
\right) \quad \mapsto \quad u_0 \sigma_0 + u_1 \sigma_1 + \cdots +u_m \sigma_m.
\end{equation}
Denote the matrices obtained from vectors $\vert v_a^s\rangle$ and $\vert w_b^t\rangle$ in this correspondence by $P_a^s$ and $Q_b^t$, respectively. Using (\ref{eq:x1/4}) it is easy to see that $P_0^s$ and $P_1^s$ are projections and due to linearity $P_0^s+P_1^s$ is equal to the corresponding matrix to the vector $\vert v_0^s\rangle + \vert v_1^s\rangle = \vert z\rangle$ which is identity. Then $\{P_0^s, P_1^s\}$, and similarly $\{Q_0^t, Q_1^t\}$ are projective measurements.

Let $\{\vert 1\rangle , \dots, \vert d\rangle\}$ be an orthonormal basis of the space on which the operators $\sigma_i$ act, and let $M^{\intercal}$ be the transpose of matrix $M$ with respect to this basis. Note that $\{Q_0^t, Q_1^t\}$ is a projective measurement, so is $\{(Q_0^t)^{\intercal}, (Q_1^t)^{\intercal}\}$. Define the bipartite vector
\begin{equation}\label{eq:epr}
\vert \psi\rangle = \frac{1}{\sqrt{d}}\sum_{i=1}^d \vert i\rangle \vert i\rangle.
\end{equation}
Now consider the following strategy for the entangled game; the shared bipartite state is $\vert \psi\rangle$, and Alice and Bob's measurement operators are $\{P_0^s, P_1^s \}$ and $\{(Q_0^t)^{\intercal}, (Q_1^t)^{\intercal}\}$, respectively. Then if Alice and Bob receive $s$ and $t$, the probability of outcomes $a$ and $b$ is equal to
\begin{eqnarray}
\langle \psi\vert P_a^s\otimes (Q_b^t)^{\intercal} \vert \psi\rangle & = & \frac{1}{d} \Tr (P_a^sQ_b^t)\nonumber\\
& = & \frac{1}{d}\left( \frac{1}{4} \Tr I+ (-1)^{a+b}\sum_{i,j=1}^m x_{s,i}y_{t,j}\Tr (\sigma_i\sigma_j) \right) \nonumber\\
& = & \frac{1}{4} + (-1)^{a+b} \sum_{i=1}^m x_{s,i}y_{t,i} \nonumber\\
&= & \langle v_a^s\vert w_b^t\rangle.
\end{eqnarray}
As a result, the winning probability of this strategy is
\begin{equation}
\sum_{s,t} \pi(s, t) \sum_{a,b} V(a, b\vert s,t)  \langle \psi\vert P_a^s \otimes (Q_b^t)^{\intercal}\vert \psi \rangle = \sum_{s,t} \pi(s, t) \sum_{a,b} V(a, b\vert s,t)  \langle v_a^s\vert w_b^t\rangle,
\end{equation}
which is equal to $\omega_{sdp}^u(G)$. We are done.

\end{proof}

The proof of this theorem indeed gives the following stronger result.

\begin{corol}\label{corol:uniform1}
$\omega_{q}(G) \geq \omega_{sdp}^u(G)$ for every binary game $G$.
\end{corol}

Note that, there exist uniform binary games which are not XOR, and this theorem cannot be directly deduced from Tsirelson's result. For a simple example of such games, assume that $S=T=\{0,1\}$, let $\pi$ be the uniform distribution over $S\times T$, and define $V(a, b\vert s,t)=1$ if and only if $a=t$ and $b=s$. This game is not XOR, but we have $\omega_{sdp}=\omega_{sdp}^u=1/2$. Then by the above Corollary $\omega_q=1/2$ and this is a uniform game. Another example is given by the uniform distribution over $\{(0, 0), (0, 1), (1, 0)\} \subset S\times T$, where $S=T=\{0,1\}$, and $V(a, b\vert s, t)=1$ iff $s\vee a \neq t\vee b$ (this game has been introduced in \cite{fl}). Again, this is not an XOR game but is uniform because $\omega_{sdp}=\omega_{sdp}^u = 2/3$. 

In both of these examples the classical value and quantum value coincide. For an example of a non-XOR uniform binary game with strictly larger entangled value one can consider a game in which with probability $1/2$ one of the above games is played and with probability $1/2$ the CHSH game.

\section{Generalizing to an arbitrary game}

We can solve the corresponding SDP of the previous section for any entangled binary game $G$ and compute its \emph{uniform} value. But a general binary game is not necessarily uniform, and then this uniform value may not be equal to $\omega_q(G)$. However, we can think of this number as an approximation of $\omega_q(G)$. Having this idea in mind, here we follow the same steps as in the previous section in order to efficiently find a lower bound on $\omega_q(G)$ for an arbitrary game $G$.

\begin{thm} \label{thm:general}
For every game $G$ we have
\begin{equation}
\omega_{sdp}(G) \geq \omega_q(G) \geq c_k\omega_{sdp}(G),
\end{equation}
where $k$ is the size of sets $A$ and $B$, and $c_k$ is given by
\begin{equation}
c_2  =  \frac{4}{3 + 2\sqrt{2}} \approx 0.68,
\end{equation}
and
\begin{equation}
c_k  =  \frac{1}{k-1},    \hspace{.3in}          \forall k\geq 3.
\end{equation}
\end{thm}

\begin{proof} The inequality $\omega_q(G) \leq \omega_{sdp}(G)$ is already discussed in Section \ref{sec:sdp}. Let $\vert z\rangle$, $\vert v_a^s\rangle$, and $\vert w_b^{t}\rangle$ denote the optimum point of the SDP of Section \ref{sec:sdp}. (Here we do not impose constraints (\ref{eq:ex1}) and (\ref{eq:ex2}).) To prove the lower bound on $\omega_q(G)$ we extract a strategy from vectors $\vert v_a^s\rangle$ and $\vert w_b^{t}\rangle $ with winning probability at least $c_k\omega_{sdp}(G)$. For simplicity, we first prove the case $k=2$ and then generalize it to an arbitrary $k$.

We can again assume that $\vert z\rangle$ is given by (\ref{eq:z}). Also suppose that the vectors $\vert v_a^s\rangle$ and $\vert w_b^{t}\rangle $ are given by (\ref{eq:vx}) and (\ref{eq:wy}).

For any $s\in S$ define the matrices
\begin{equation}
M_0^s =  x_{s,1}\sigma_1 +\cdots + x_{s, m}\sigma_m,
\end{equation}
and $M_1^s = - M_0^s$. It is easy to see that
\begin{equation}
(M_0^s)^2 = (M_1^s)^2 = \left(\sum_{i=1}^m x_{s,i}^2 \right) I.
\end{equation}
So the eigenvalues of both $M_0^s$ and $M_1^s$ are $\pm (\sum_{i= 1}^m x_{s,i}^2)^{1/2}$.
In the previous section we had $x_{s,0}=1/2$ and $\sum_{i=1}^m x_{s,i}^2 = 1/4$, and concluded that the matrices $(1/2) I + M_0^s$ and $(1/2) I + M_1^s$ are projections which can be considered as the measurement operators of Alice. However, we do not really require Alice's operators to be projections; instead, we need them to be positive semidefinite and sum to identity (to consist POVM measurements). So we slightly change the matrices $x_{s,0}I+ M_0^s$ and $(1-x_{s,0}) I + M_1^s$ in order to obtain positive semidefinite matrices.

Assume that $x_{s,0}\geq 1/2$ (the other case $1-x_{s,0} >1/2$ is similar).\footnote{Notice that $x_{s,0} = \langle z\vert v_0^s\rangle = \langle v_0^s\vert v_0^s\rangle$, and thus $0\leq x_{s,0}\leq 1$.} Since $\vert v_0^s\rangle$ and $\vert v_1^s\rangle$ are orthogonal and
\begin{equation}
\langle v_0^s\vert v_1^s\rangle = x_{s,0}(1-x_{s,0}) - \sum_{i=1}^{m} x_{s,i}^2
\end{equation}
we have $\sum_{i=1}^m x_{s,i}^2=x_{s,0}(1-x_{s,0}) \leq x_{s,0}^2$. This means that the eigenvalues of both matrices
$ x_{s,0}I + M_{0}^s$ and $\sqrt{x_{s,0}(1-x_{s,0})}I + M_{1}^s$ are non-negative. Therefore,
\begin{equation}
P_0^s = \frac{1}{x_{s,0}+\sqrt{x_{s,0}(1-x_{s,0})}} \left( x_{s,0}I + M_{0}^s \right)
\end{equation}
and
\begin{equation}
P_1^s = \frac{1}{x_{s,0}+\sqrt{x_{s,0}(1-x_{s,0})}} \left( \sqrt{x_{s,0}(1-x_{s,0})}I + M_{1}^s \right)
\end{equation}
are positive semidefinite, and also since $M_1^s= - M_0^s$, $P_0^s+P_1^s = I$. Similarly positive semidefinite matrices $Q_0^t, Q_1^t$ which sum to identity can be defined in terms of vectors $\vert w_0^t\rangle$ and $\vert w_1^t\rangle$.

Now consider the strategy given by the bipartite state $\vert \psi\rangle$ given by (\ref{eq:epr}) and POVM measurements $\{P_0^s, P_1^s\}$ and $\{(Q_0^t)^{\intercal}, (Q_1^t)^{\intercal}\}$. It is easy to see that for any $s, t, a$ and $b$ we have
\begin{eqnarray}
\langle \psi\vert P_a^s\otimes (Q_b^t)^{\intercal} \vert \psi\rangle & = & \frac{1}{d} \Tr (P_a^s Q_b^t) \nonumber\\
& \geq  & \frac{1}{\alpha_{s} + \sqrt{x_{s,0}(1-x_{s,0})}} \cdot \frac{1}{\beta_{t} + \sqrt{y_{t,0}(1-y_{t,0})}} \langle v_a^s\vert w_b^t\rangle,
\end{eqnarray}
where $\alpha_{s} = \max \{x_{s,0}, 1-x_{s,0}\}$ and $\beta_t= \max \{y_{t,0}, 1-y_{t,0}\}$. Also, observe that the minimum of
\begin{equation}
\frac{1}{ \alpha + \sqrt{\alpha (1-\alpha)} }
\end{equation}
in the interval $[1/2, 1]$ is equal to $2/(1+\sqrt{2})$. Thus we conclude that
\begin{equation}
\langle \psi\vert P_a^s\otimes (Q_b^t)^{\intercal} \vert \psi\rangle \geq \left( \frac{2}{1+\sqrt{2}} \right)^2 \langle v_a^s\vert w_b^t\rangle.
\end{equation}
As a result, the winning probability of this strategy
\begin{equation}\sum_{s,t} \pi(s,t) \sum_{a,b} V(a, b\vert s,t) \langle \psi\vert P_a^s\otimes (Q_b^t)^{\intercal} \vert \psi\rangle
\end{equation}
is at least
\begin{eqnarray}
\left(\frac{2}{1+\sqrt{2}} \right)^2\sum_{s,t} \pi(s,t) \sum_{a,b} V(a, b\vert s,t) \langle v_a^s\vert w_b^t\rangle = \frac{4}{3 + 2\sqrt{2}}\, \omega_{sdp}(G).
\end{eqnarray}
Therefore, $\omega_q(G)\geq \frac{4}{3 + 2\sqrt{2}}\, \omega_{sdp}(G) $.

The proof for a larger $k$ is similar to $k=2$. Again, to every vector $\vert v_a^s\rangle$ and $\vert w_b^t\rangle$ we correspond a positive semidefinite matrix as follows. Let $\vert u^{(1)}\rangle, \dots , \vert u^{(k)}\rangle$ be the vectors corresponding to either a question $s$ from Alice, or question $t$ from Bob. Then $\langle u^{(i)}\vert u^{(j)} \rangle =0 $ for any $i\neq j$ and $\sum_i \vert u^{(i)} \rangle = \vert z\rangle$. Let
\begin{equation}
\vert u^{(i)}\rangle =\left(
  \begin{array}{c}
    u_0^{(i)} \\
    u_1^{(i)} \\
    \vdots \\
    u_m^{(i)} \\
  \end{array}
\right).
\end{equation}
Because $ \langle u^{(i)} \vert u^{(i)}\rangle= \langle z\vert u^{(i)}\rangle = u_0^{(i)}$, the matrix
\begin{equation}
\sqrt{u_0^{(i)}(1-u_0^{(i)})} I + u_1^{(i)} \sigma_1 +\cdots + u_m^{(i)} \sigma_m
 \end{equation}
is positive semidefinite. So we correspond the following POVM elements to the vectors $\vert u^{(1)}\rangle, \dots , \vert u^{(k)}\rangle$
\begin{equation}
\frac{1}{\sum_{j=1}^k\, \max \{u_0^{(j)}, \sqrt{u_0^{(j)}(1-u_0^{(i)})} \,\}} \left( \max \{ u_0^{(i)}, \sqrt{u_0^{(i)}(1-u_0^{(i)})} \} I + u_1^{(i)} \sigma_1 +\cdots + u_m^{(i)} \sigma_m  \right).
\end{equation}
Therefore, considering these measurement operators (again, Bob's operators are replaced by the transpose of these matrices) and the state $\vert \psi\rangle$ defined by (\ref{eq:epr}), we obtain $\omega_q(G) \geq c_k\omega_{sdp}(G)$. The constant $c_k$ is given by
\begin{equation}
c_k =  \min \left( \frac{1}{ \max \{ u_0^{(1)} , \sqrt{ u_0^{(1)}(1-u_0^{(1)})}  \} + \cdots +\max \{ u_0^{(k)} , \sqrt{ u_0^{(k)}(1-u_0^{(k)})}  \}   } \right)^2,
\end{equation}
where the minimum is taken over all non-negative numbers $ u_0^{(1)}, \dots, u_0^{(k)}$ that sum to one.\footnote{ They sum to one because $u_0^{(1)} +\cdots +u_0^{(k)} = \langle z\vert u^{(0)}\rangle + \cdots + \langle z\vert u^{(k)}\rangle = \langle z\vert z\rangle =1. $ } To compute the minimum one can simplify the expression by considering the following two cases and use the Lagrange multiplier method: first, for all $i$, $u_0^{(i)}\leq 1/2$ which means that $\max \{ u_0^{(i)} , \sqrt{u_0^{(i)}(1-u_0^{(i)})}\} = \sqrt{u_0^{(i)}(1-u_0^{(i)})} $, and second, $u_0^{(1)}> 1/2$ which implies $u_0^{(i)}\leq 1/2$ for all $i>1$. In both cases, by \emph{fixing} $u_0^{(1)}$, the problem can be reformulated as maximizing 
\begin{align}
\sqrt{u_0^{(2)}(1-u_0^{(2)})} + \cdots + \sqrt{u_0^{(k)}(1-u_0^{(k)})},
\end{align}
conditioned on $ 0\leq u_0^{(2)}, \dots, u_0^{(k)} \leq 1/2 $, and $u_0^{(2)}+ \cdots+ u_0^{(k)} = 1 - f$, where in the first case $f = \sqrt{u_0^{(1)}(1-u_0^{(1)})}$, and in the second case $f= u_0^{(1)}$. Then the Lagrange multiplier condition gives $u_0^{(2)}= \dots =u_0^{(k)} = (1-f)/(k-1)$. On the other hand, at the boundary of feasible points where one of $u_0^{(i)}$, $2\leq i\leq k$, is equal to $0$ or $1/2$, the problem essentially reduces to the same problem but with a smaller $k$, on which we can again apply the above Lagrange multiplier condition. Comparing these different cases we find that the optimum point for $k\geq 3$ is $u_0^{(1)}= \dots =u_0^{(k)} = 1/k$ and $c_k = 1/(k-1)$.
\end{proof}

\section{Discussion}

In this paper, based on the work of Tsirelson \cite{Tsirelson}, we introduce a way of embedding a set of given vectors into the cone of positive semidefinite matrices such that the inner product is approximately preserved. Any improvement in this direction would imply stronger bounds on the entangled value of games based on SDP relaxation. Here we should mention that the ``rounding" algorithm of Kempe, Regev, and Toner \cite{unique} can also be seen as such embedding which works for every game and not only unique ones. So it is a valid question whether their rounding method appled to games with small $k$ (say binary games) improves our results.

Approximation algorithms based on SDP relaxation have also been applied to compute the classical value of games, specially the binary \emph{constraint satisfaction problem} (2-CSP) which is defined as follows. We are given $n$ Boolean variables 
$x_1, \dots, x_n$ and a constraint (maybe trivial) on any pair $x_i, x_j$ with a weight $w_{ij}\geq 0$, and the problem is find an assignment of the variables that maximizes the summation of the weights of satisfied pairs. It is easy to see that 2-CSP problem is equivalent to MAX~DI-CUT which is the following problem. Consider a directed graph on the vertex set $\{1, \dots, n\}$ with a weight $w_{ij}\geq 0$ for every directed edge $i\rightarrow j$. The goal is to find a partition $\{1, \dots, n\} = V_1\cup V_2$ that maximizes the total weight of edges $i\rightarrow j$ where $i\in V_1$ and $j\in V_2$. In~\cite{2-csp} an SDP relaxation is corresponded to MAX~DI-CUT which after a change of variables can be written as follows.
\begin{alignat}{2}
    \text{Maximize:} \quad & \sum_{i, j=1}^{n}  w_{ij} \langle v_i\vert v_{n+j}\rangle \nonumber \\
    \text{Subject to:} \quad
    & \langle z\vert z\rangle =1, \nonumber\\
    & \vert v_i\rangle + \vert v_{n+i}\rangle = \vert z\rangle, \quad & i=1,\dots ,n\nonumber\\
     & \langle v_i\vert v_{n+i}\rangle =0, \quad & i=1,\dots, n \nonumber\\
    & \langle v_i\vert v_j\rangle \geq 0, \quad & i,j=1, \dots, 2n.   \label{eq:sdp-23}
\end{alignat}
This SDP is similar to the SDP~\eqref{eq:sdp} for binary games; one difference is that~\eqref{eq:sdp} is \emph{bipartite}. It is proved in~\cite{2-csp} that the above SDP relaxation gives a $0.874$-approximation algorithm for MAX DI-CUT. 
Now the question is whether this algorithm, with such accuracy, can be generalized to binary games or not.

Finding examples to examine the efficiency of the bounds $c_2\approx 0.68$ and $c_k=1/(k-1)$ for $k\geq 3$, is of interest. Of course, the main difficulty is estimating $\omega_q$ for large size games. Curiously, for all known examples of binary games to the author with $\vert S\vert =\vert T\vert =2$ (as those in Section \ref{sec:uniform}) $\omega_q$ is either equal to $\omega_c$ or $\omega_{sdp}^u$; the reason is that in all such examples the shared state between players is a maximally entangled state. For the proof of the following proposition we use similar ideas as in \cite{masanes}.

\begin{proposition}
Suppose that Alice and Bob are restricted to share a maximally entangled state of an arbitrary dimension. Then for every binary game $G$ with $S=T=\{0,1\}$, $\omega_q(G)$ is either equal to $\omega_c(G)$ or $\omega_{sdp}^u(G)$.
\end{proposition}

\begin{proof}
Using Proposition 2 of \cite{CHTW} we may assume that Alice and Bob's measurements are projective: $\{P^s_0, P^s_1\}$ and $\{Q^t_0, Q^t_1\}$, $s,t\in \{0,1\}$. Note that two projections $P^0_0, P_0^1$, and then $P^0_1=I - P^0_0, P^1_1 = I -P^1_0$, can simultaneously be block--diagonalized with blocks of size at most $2$. Letting $S_1, \dots, S_m$ be projections on the corresponding 1 or 2-dimensional subspaces, we conclude that the protocol is equivalent to one in which Alice before receiving $s$ measures her part of the shared state using the projective measurement $\{S_1, \dots, S_m\}$ and then proceeds as before. On the other hand, when we project a maximally entangled state onto a subspace on the Alice side, Bob's part of the state would be projected onto the similar subspace, and the resulting state would again be a maximally entangled state. Thus we may also assume that Bob before receiving $t$ measures his part of the shared state using the projective measurement $\{S_1, \dots, S_m\}$. As a result, since these two projective measurements are preformed before receiving $s, t$ and their outcomes coincide, Alice and Bob may start by sharing the post-measurement states $\vert \psi_1\rangle, \dots, \vert \psi_m\rangle$ (where $\vert \psi_i\rangle$ is the maximally entangled state defined on $S_i\otimes S_i$) and by sharing randomness choose $i$, and then proceed through the rest of the protocol by measuring $\vert \psi_i\rangle$ using POVMs $\{P^s_0, P^s_1\}$ and $\{Q^t_0, Q^t_1\}$. Now note that shared randomness never helps Alice and Bob to increase their winning probability. So they obtain the optimal strategy by sharing only one of the states $\vert \psi_1\rangle, \dots , \vert \psi_m\rangle$ and we may assume that the shared state is the maximally entangled state of dimension $2\times 2$.

Suppose that one of Alice's operators $P_a^s$, is zero. This means that all projections $P_0^0, P_1^0, P_0^1, P_1^1$ on the Alice side commute, and then the whole strategy is a classical one and $\omega_q(G)=\omega_c(G)$. On the other hand, if all $P_a^s$ and $Q_b^t$ are non-zero we obtain $\Tr(P_a^s)=\Tr(Q_b^t)=1$, which because the shared state is the maximally entangled state, means that the strategy is uniform. Then by Theorem \ref{thm:uniform} we have $\omega_q(G)=\omega_{sdp}^u(G)$.

\end{proof}

Here we should mentioned that there are binary games whose entangled value is strictly between its classical value and the optimum value of the corresponding SDP relaxation. For example the I3322 inequality can be modified to have positive coefficients and be a game with the above property, see \cite{dltw}.

Another observation which may shed a light on this problem is a simple generalization of the result of \cite{CHTW} that for a binary game $G$ if $\omega_q(G)=1$ then $\omega_{c}(G)=1$.

\begin{proposition} Let $G$ be a binary game. If $\omega_{sdp}(G)=1$, then $\omega_q(G)=\omega_c(G)=1$.
\end{proposition}

\begin{proof} Let $\vert v_a^s\rangle, \vert w_b^t\rangle$, and $\vert z\rangle$ be the optimal vectors in the SDP relaxation associated to $G$. Then since $\omega_{sdp}(G)=1$, for every $a, b, s,t$ such that $\pi(s,t)\neq 0$ and $V(a, b\vert s, t)=0$ we have $\langle v_a^s\vert w_b^t\rangle =0$. Without loss of generality, we may assume that the converse also holds ($V(a, b\vert s, t)=1$ if and only if $\langle v_a^s\vert w_b^t\rangle \neq 0$) because by changing $V(a, b\vert s, t)$ from $1$ to $0$ we only decrease the value of the game.

Define a bipartite graph on the vertex set $S\cup T$ as follows. Connect vertices $s\in S$ and $t\in T$ if $\pi(s, t)\neq 0$ and the set of pairs $(a, b)$ such that $V(a, b\vert s, t)=1$ is equal to either $\{(0,0),(1, 1)\}$ or $\{(0,1),(1,0)\}$. Assume that we are in the first case; that is, $V(a, b\vert s, t)=1$ if and only if $a=b$. We conclude that $\langle v_0^s\vert w_1^t\rangle = \langle v_1^s\vert w_0^t\rangle =0$, and then
\begin{align}
\langle v_0^s\vert v_0^s\rangle = \langle v_0^s\vert z\rangle = \langle v_0^s\vert w_0^t\rangle = \langle z\vert w_0^t\rangle =\langle w_0^t\vert w_0^t\rangle  .
\end{align}
Thus by the Cauchy--Schwarz inequality $\vert v_0^s\rangle = \vert w_0^t\rangle$, and similarly $\vert v_1^s\rangle = \vert w_1^t\rangle$. As a result, for every connected component of the bipartite graph there are two orthogonal vectors $\vert u_0\rangle$ and $\vert u_1\rangle= \vert z\rangle - \vert u_0\rangle$ such that for every vertex $s$ or $t$ in this component we have $\{ \vert v_0^s\rangle, \vert v_1^s\rangle\} = \{ \vert w_0^t\rangle, \vert w_1^t\rangle\}=\{ \vert u_0\rangle, \vert u_1\rangle\}$. 

Now consider the following \emph{classical} strategy. In each connected component, between the corresponding vectors $\vert u_0\rangle$ and $\vert u_1\rangle$ pick the longer one (if they have equal lengths pick one of them arbitrarily); then Alice upon receiving $s$ outputs $a$ such that $\vert v_a^s\rangle $ is equal to the chosen vector in the corresponding component. Bob's strategy is defined similarly. We prove that the winning probability of this strategy is $1$. 

Fix $s, t$ such that $\pi(s,t)\neq 0$. We show that if Alice and Bob's questions are $s$ and $t$, their answers $a, b$ satisfy $V(a, b \vert s, t)=1$.  By changing $0$ to $1$ or vice versa, we may assume that the set of pairs $(a, b)$ such that $V(a, b \vert s, t)=1$ has one of the following forms. In each case we show that Alice and Bob always win.
\begin{itemize}
\item $\{0,1\}\times \{0, 1\}$: no matter what their outcomes are, Alice and Bob always win in this case.

\item $\{(0,0),(0,1),(1,0)\}$: note that in this case $\langle v_a^s\vert w_b^t\rangle \neq 0$ if and only if $(a, b)\neq (1, 1)$. Then it is easy to see that $\langle v_1^s\vert v_1^s \rangle < \langle w_0^t\vert w_0^t\rangle$ and $\langle w_1^t\vert w_1^t \rangle < \langle v_0^s\vert v_0^s\rangle$. Therefore, in the classical strategy $1$ cannot be the simultaneous answer of both Alice and Bob.

\item $\{(0, 0), (1, 1)\}$: the whole strategy is indeed defined based on this case; $s$ and $t$ belong to the same connected component of the graph and if Alice outputs $0$, Bob's output is also $0$ because in this case $\vert v_0^s\rangle =\vert w_0^t\rangle $ and $\vert v_1^s\rangle = \vert w_1^t\rangle$.

\item $\{(0,0),(0,1)\}$: we have $\langle v_1^s \vert v_1^s\rangle= \langle v_1^s\vert z\rangle = \langle v_1^s\vert w_0^t\rangle  + \langle v_1^s\vert w_1^t\rangle =0$. So $\vert v_1^s\rangle$ is shorter than $\vert v_0^s\rangle$ and Alice's output is $0$.

\item $\{(0,0)\}$: similar to the previous case both $\vert v_1^s\rangle$ and $\vert w_1^t\rangle$ would be zero.
\end{itemize}
We are done.

\end{proof}

This proposition \emph{suggests} that if for a binary game $\omega_{sdp}(G)$ is close to $1$, then $\omega_{q}(G)$ (or even $\omega_c(G)$) is also close to $1$. For the special case of 2-CSP it is proved in~\cite{near-optimal} that $\omega_{sdp} = 1-\epsilon$ implies $\omega_c= 1- O(\sqrt{\epsilon})$. 

Another problem regarding estimation of the entangled value of games is to find rounding algorithms corresponding to the higher levels of the SDP hierarchy proposed in \cite{dltw, npa}.

\vspace{.3in}

\noindent{\bf Acknowledgements.} The author is grateful to Leonard Schulman and Stephanie Wehner for several useful discussions, and to unknown referees whose comments significantly improved the presentation of the paper. This work is partially supported by NSF under Grant No. PHY-0803371 and by NSA/ARO under Grant No. W911NF-09-1-0442.


\small
\begin{thebibliography}{10}
\vspace*{.04in}


\bibitem{CGJ} R. Cleve, D. Gavinsky, and R. Jain. Entanglement-resistant two-prover interactive proof systems and non-adaptive private information retrieval systems. {\it Quantum Information and Computation}, Vol.9 No.7\&8, 648--656, 2009.

\bibitem{CHTW} R. Cleve, P. H{\o}yer, B. Toner, and J. Watrous. Consequences and limits of nonlocal strategies. {\it  In Proceedings of
the 19th Annual IEEE Conference on Computational Complexity}, 236--249, 2004.

\bibitem{dltw} A. Doherty, Y-C Liang, B. Toner, and S. Wehner. The quantum
moment problem and bounds on entangled multi-prover games. {\it In Proceedings of
the 23rd Annual IEEE Conference on Computational Complexity}, 199-210, 2008.


\bibitem{grothen} J. Bri{\"e}t, H. Buhrman, and B. Toner. A generalized Grothendieck inequality and entanglement in XOR games, {\it The 12th Workshop on Quantum Information Processing}, 2009,  arXiv:0901.2009.

\bibitem{vertesi} T. V\'ertesi and K.F. P\'al. Bounding the dimension of bipartite quantum systems. {\it Physical Review A} 79, 42106, 2009.

\bibitem{fl} U. Feige and L. Lov\'{a}sz. Two-Prover One-Round Proof Systems:
Their Power and Their Problems. {\it In Proceedings of the 24th Annual ACM Symposium on Theory of Computing}, 733--744, 1992.


\bibitem{Hastad} J.~H{\aa}stad. Some optimal inapproximability results.
{\it Journal of the ACM}, 48(4):798--859, 2001.

\bibitem{Tsirelson} B.~S. Tsirelson. Quantum generalizations of Bell's inequality. {\it Letters in Mathematical Physics}, 4(2):93--100, 1980.

\bibitem{ito} T. Ito, H. Kobayashi, and K. Matsumoto. Oracularization and two-prover one-round interactive proofs against nonlocal strategies. {\it In Proceedings of the 24th Annual IEEE Conference on Compuational Complexity}, 217--228, 2009.

\bibitem{qip-pspace} R. Jain, Z. Ji, S. Upadhyay, and J. Watrous. QIP = PSPACE. {\it In Proceedings of the 42nd Annual ACM Symposium on Theory of Computing}, to appear, arXiv:0907.4737.

\bibitem{qip2-pspace} R. Jain, S. Upadhyay, and J. Watrous. Two-message quantum interactive proofs are in PSPACE. arXiv:0905.1300.


\bibitem{20} J. Kempe, H. Kobayashi, K. Matsumoto, B. Toner, and T. Vidick. Entangled Games are Hard to Approximate. {\it In Proceedings of the 49th Annual Symposium on Foundations of Computer Science}, 447--456, 2008.

\bibitem{unique} J. Kempe, O. Regev, and B. Toner. Unique Games with Entangled Provers are Easy. {\it In Proceedings of the 49th Annual Symposium on Foundations of Computer Science}, 457--466, 2008.

\bibitem{npa} M. Navascu\'es, S. Pironio, and A. Ac\'\i
n. A convergent hierarchy of semidefinite
programs characterizing the set of quantum correlations. {\it New Journal of Physics},
10:073013, 2008.



\bibitem{19} T. Ito, H. Kobayashi, D. Preda, X. Sun, and A. C.-C Yao. Generalized Tsirelson Inequalities, Commuting-Operator Provers, and Multi-prover Interactive Proof Systems. {\it In Proceedings of the 23rd Annual IEEE Conference on Computational Complexity}, 187--198, 2008.



\bibitem{ste} S. Wehner. Entanglement in Interactive Proof Systems with Binary Answers. {\it In Proceedings of the 23rd Annual Symposium on Theoretical Aspects of Computer Science}, 162--171, 2006




\bibitem{2-csp} M. Lewin, D. Livnat, and U. Zwick. Improved Rounding Techniques for the MAX 2-SAT and MAX DI-CUT Problems. {\it In Proceedings of the 9th International Conference on Integer Programming and Combinatorial Optimization}, 67--82, 2002.



\bibitem{masanes} L. Masanes. Extremal quantum correlations for N parties with two dichotomic observables per site. quant-ph/0512100, 2005.



\bibitem{near-optimal} M. Charikar, K. Makarychev, and Y. Makarychev. Near-optimal algorithms for maximum constraint satisfaction problems. {\it In Proceedings of the 18th Annual ACM-SIAM Symposium on Discrete Algorithms}, 62--68, 2007.



\end{thebibliography}
\end{document}